\documentstyle[aps, epsf, psfig]{revtex}

\newcommand{\vphi}{\varphi}

\newcommand{\pa}{\partial}

\newcommand{\sla}[1]{\slash\!\!\! ~#1}
\newcommand{\td}{\tilde}
\newcommand{\ep}{\epsilon}
\begin{document}

\draft
\begin{center}
{\Large\bf Light Quark Masses to All Order of Chiral Expansion}\\[5mm]
Xiao-Jun Wang\footnote{E-mail address: wangxj@mail.ustc.edu.cn} \\
{\small
Center for Fundamental Physics,
University of Science and Technology of China\\
Hefei, Anhui 230026, P.R. China} \\
Mu-Lin Yan\\
{\small CCST(World Lad), P.O. Box 8730, Beijing, 100080, P.R. China}\\
{\small and}\\
{\small  Center for Fundamental Physics,
University of Science and Technology of China\\
Hefei, Anhui 230026, P.R. China}\footnote{mail
address}
\end{center}
\begin{abstract}
\noindent
We study light (current) quark masses in framework of chiral constitutent
quark model.  In our calculation the current quark masses are defined
uniquely, and all order effects of the light quark masses are considered.
The results at energy scale $\mu=m_\rho$ are $m_s=(160\pm 15)$MeV,
$m_s/m_d=20.2\pm 3.0$, $m_u/m_d=0.5\pm 0.09$. 
\end{abstract}

\section{Introduction}

The quark masses are some of basic parameters of the standard model. There
are various hadronic phenomenologies relating to the light
quark($u,\;d,\;s$) masses. For instance, they break the chiral symmetry of
QCD explicitly. $m_u-m_d$ breaks isospin symmetry or charge symmetry, and
$(m_u+m_d)/2-m_s$ breaks SU(3) symmetry in hadron physics respectively.
However, in QCD the masses of the light quarks are not directly
measureable in inertial experiments, but enter the theory only indrectly
as parameters in the fundamental lagrangian. The purpose of this paper is
to study light quark masses at energy scale $\mu=m_\rho$ in a
non-perturbative way.

In general, at low energy information about the light quark mass ratios is
extracted in order by order by a rigorous, semiphenomenological method,
chiral perturbative theory(ChPT)\cite{Wein79,GL82,GL85a}. To the first
order, the results, $m_s/m_d=19$ and $m_u/m_d=0.556$, are
well-known\cite{GMO68}. 
Many authors have studied the mass ratios up to next to leading order
of the chiral expansion. Gasser and Leutwyler first obtained
$m_s/m_d=20.2$ and $m_u/m_d=0.554$\cite{GL85a}. Then Kaplan and Manohar
extracted $m_s/m_d=$15 to 23 and $m_u/m_d=$0 to 0.8 with very larger error
bar\cite{Kaplan86}. These values have been improved to $m_s/m_d=(20.5\pm
2.5),\;m_u/m_d=0.52\pm 0.13$ by Leutwyler\cite{Leutwyler90}, to
$m_s/m_d=18,\;m_u/m_d=0.66$ by Gerard\cite{Gerard90}, and to
$m_s/m_d=21,\;m_u/m_d=0.30\pm 0.07$ by Donoghue {\sl et.
al.}\cite{Donoghue92} respectively. Finally, Leutwyler analysis previous
results and
obtained $m_s/m_d=18.9\pm 0.8,\;m_u/m_d=0.553\pm 0.043$\cite{Leut96}. So
far, however, the study on light quark masses in framework of ChPT is
limited by the following shortages: 1) In this framework, to obtain light
quark mass ratio beyond the next to leading-order is very diffcult, since
more and more free parameters are included with raising of perturbative
order. 2) Due to Kaplan-Manohar ambiguity\cite{Kaplan86}, the
defination of the light quark masses in ChPT is not uniquely. The reason
has been pointed out in ref.\cite{Donoghue92} that, at the next to
leading-order, QCD renormalization is mass-dependent, and the symmetry can
not distinguish renormalized quark masses from ``bare'' quark masses. 
3) In framework of ChPT, we can only obtain light
quark masse ratios. For obtaining the individual quark masses, other
approachs, such as QCD sum rules\cite{Sum1,Sum2} or lattice
calculation\cite{Lattice} are needed. These shortages are urgently wanted
to be improved by theorectical studies of QCD and experiment.

In ref.\cite{Rho} we have the constructed chiral constituent quark
model(ChCQM) including the lowest vector meson resonances following the
spirit of Manohar-Georgi model\cite{MG84}. This model provides a
formulation to perform rigorous field theory calculation at energy scale
lying between
ChPT($\mu\sim 0.5$GeV) and chiral symmetry spontaneously broken(CSSB)
scale($\mu\sim 1.2$GeV), and a successful description on physics in this
energy region\cite{Rho,Omega}. Up to $O(p^4)$, low energy limit of the
model agree with ChPT well. Thus this model can be treated as an approach
to extend ChPT investigation inspired by QCD. The most important advantage
of this approach is that we can perform calculation beyond the low energy
expansion in ChCQM, and only fewer free parameters are required. In
general, there are three types of expansion working at low energy. They
are momentum expansion, light quark mass expansion and $N_c^{-1}$
expansion\cite{tH74}. In ref.\cite{Rho,Omega}, we have provided a rigorous
method to perform calculation to include all order terms of the momentum
expansion and up to the next to leading order of $N_c^{-1}$ expansion. In
this the present paper, we will
extend this method to reflect all order information of light quark mass
expansion.

The Kaplan-Manohar ambiguity of ChPT has caused many debates. In
particular, due to this ambiguity, the authors of ref\cite{Choi88} argued
that the observed mass spectrum is consistent with a broad range of quark
mass ratios, which specially includes the possibility $m_u=0$. However, 
it is disagreed by other anthors\cite{Leut90,Donoghue92}. This problem can
also be discussed in ChCQM. Since ChCQM is an effective approach with
features of low energy QCD, light current quark masses are defined
uniquely in ChCQM, that are just renormalized ``physical'' masses of u, d,
and s quarks. In principle, therefore, there is no Kaplan-Manohar
ambiguity in ChCQM. Light current quark masses can be determined uniquely
via meson spectrum. 

In ref.\cite{Omega}, we studied the isospin breaking process
$\omega\rightarrow\pi^+\pi^-$ and determined isospin breaking parameter
$m_d-m_u=(3.9\pm 0.22)$MeV at vector meson energy scale. This result
together with meson spectrum provide so much information that we can
obtain not only light quark mass ratios but also individual quark masses. 

In ref.\cite{Rho}, we have shown that the chiral
expansion at vector meson energy scale converge slowly. In particular, we
have also pointed out that, if we neglect strange quark masses, the chiral
expansion at $\phi(1020)$ energy scale will be divergent! It implies that
$m_s$ play very important role at $\phi$-physics. In ref.\cite{Rho,Omega}
we have successfully studies the chiral expansion at $m_\rho$ and
$m_\omega$ energy scale. In order to extend this study to $K^*(892)$ and
$\phi(1020)$, individual quark masses are neccessary. It has been
recognized that the light quark masses obtained in different approaches
are with larger difference. Thus we have to extract quark masses by
this formalism itself. It is one of our goals.

In general, in ChCQM the light quark masses can be extracted not only by
pseudoscalar meson spectrums, but also by the lowest vector meson
resonance spectrums. However, one-loop effects of mesons will contribute
to vector meson masses, and calculation on one-loop effects of mesons
is related to the chiral expansion at vector meson energy scale. As shown
in ref.\cite{Rho}, this relation is very complicate. It makes that the
relationship between vector meson spectrums and the light quark masses are
also very complicate and indirect. Thus in this paper we will extract
information about the light current quark masses from pseudoscalar meson
spectrums and their decay constants. The vector meson spectrums will be
predicted by this formalism in other paper.

The contents of the paper are organized as follows. In sect. 2 we review
the basic notations of the chiral constituent quark model with the lowest
vector meson resonances. In sect. 3, masses and decay constants of
pseudoscalar meson octet are calculated. The results will including all
order information of the light quark masses. In sect. 4, one-loop effects
of pseudoscalar mesons and renormalization are discussed. The numerical
results are given in sect. 5 and a brief summary is included in sect. 6.

\section{Chiral Constituent Quark Model}

The simplest version of chiral quark model which was originated by
Weinberg\cite{Wein79}, and developed by Manohar and Georgi\cite{MG84}
provides a QCD-inspired description on the simple constituent quark model.
In view of this model, in the energy region between the CSSB scale and the
confinement scale ($\Lambda_{QCD}\sim 0.1-0.3 GeV$), the dynamical field
degrees of freedom are constituent quarks(quasi-particle of quarks),
gluons and Goldstone bosons associated with CSSB(these Goldstone bosons
correspond to lowest pseudoscalar octet). In this quasiparticle
description, the effective coupling between gluon and quarks is small and
the important interaction is the coupling between quarks and Goldstone 
bosons. In {\bf I} we have further included the lowest
vector meson resonances into this formalism. At chiral limit, this model
is
parameterized by the following chiral constituent quark lagrangian
\begin{eqnarray}\label{2.1}
{\cal L}_{\chi}&=&i\bar{q}(\sla{\pa}+\sla{\Gamma}+
  g_{_A}{\slash\!\!\!\!\Delta}\gamma_5-i\sla{V})q-m\bar{q}q
   +\frac{F^2}{16}<\nabla_\mu U\nabla^\mu U^{\dag}>
   +\frac{1}{4}m_0^2<V_\mu V^{\mu}>.
\end{eqnarray}
Here $<...>$ denotes trace in SU(3) flavour space,
$\bar{q}=(\bar{q}_u,\bar{q}_d,\bar{q}_s)$ are constituent quark fields.
$V_\mu$ denotes vector meson octet and singlet. Since in this paper we
only focus on pseudoscalar meson spectrums and decay constants, we will
neglect vector meson fields in the following. The $\Delta_\mu$ and
$\Gamma_\mu$ are defined as follows,
\begin{eqnarray}\label{2.2}
\Delta_\mu&=&\frac{1}{2}\{\xi^{\dag}(\pa_\mu-ir_\mu)\xi
          -\xi(\pa_\mu-il_\mu)\xi^{\dag}\}, \nonumber \\
\Gamma_\mu&=&\frac{1}{2}\{\xi^{\dag}(\pa_\mu-ir_\mu)\xi
          +\xi(\pa_\mu-il_\mu)\xi^{\dag}\},
\end{eqnarray}
and covariant derivative are defined as follows
\begin{eqnarray}\label{2.3}
\nabla_\mu U&=&\pa_\mu U-ir_\mu U+iUl_\mu=2\xi\Delta_\mu\xi,
  \nonumber \\
\nabla_\mu U^{\dag}&=&\pa_\mu U^{\dag}-il_\mu U^{\dag}+iU^{\dag}r_\mu
  =-2\xi^{\dag}\Delta_\mu\xi^{\dag},
\end{eqnarray}
where $l_\mu=v_\mu+a_\mu$ and $r_\mu=v_\mu-a_\mu$ are linear combinations
of external vector field $v_\mu$ and axial-vector field $a_\mu$, $\xi$
associates with non-linear realization of spontanoeusly broken global
chiral symmetry introduced by Weinberg\cite{Wein68}. This realization is
obtained by specifying the action of global chiral group $G=SU(3)_L\times
SU(3)_R$ on element $\xi(\Phi)$ of the coset space $G/SU(3)_{_V}$:
\begin{equation}\label{2.4}
\xi(\Phi)\rightarrow
g_R\xi(\Phi)h^{\dag}(\Phi)=h(\Phi)\xi(\Phi)g_L^{\dag},\hspace{0.5in}
 g_L, g_R\in G,\;\;h(\Phi)\in H=SU(3)_{_V}.
\end{equation}
Explicit form of $\xi(\Phi)$ is usual taken
\begin{equation}\label{2.5}
\xi(\Phi)=\exp{\{i\lambda^a \Phi^a(x)/2\}},\hspace{1in}
U(\Phi)=\xi^2(\Phi),
\end{equation}
where the Goldstone boson $\Phi^a$ are treated as pseudoscalar meson
octet. In ref.\cite{Rho} we have shown that the lagrangian(~\ref{2.1}) is
invariant under $G_{\rm global}\times G_{\rm local}$.
 
Distingushing from some versions of chiral quark models, there is a
kinetic term of pseudoscalar mesons in lagrangian(~\ref{2.1}). Therefore,
the kinetic term of pseudoscalar mesons generated by one-loop effects of
constituent quarks can be renormalized. Note that there is no mass term of
pseudoscalar mesons in eq.~(\ref{2.1}).  All of these are due to the
basic assumption of the model\cite{MG84}. In the other words, in this
energy region, the dynamical field degrees of freedom are constituent
quarks and massless Goldstone bosons(pseudoscalar octet) associated with
CSSB. Masses of pseudoscalar mesons will be genarated by quark loops as
current quark mass parameters emerge in the dynamics (to see below). In
ref.\cite{Rho}
we have fitted the parameter $g_A=0.75$ via $\beta$-decay of neutron, and
$m=480$MeV via low energy limit of the model. It has been also pointed out
that the value of $g_A$ has included effects of intermediate axial-vector
meson resonances exchanges at low energy.

The light current quark mass-dependent term has been introduced in
ref.\cite{Omega} based on requirement of the chiral symmetry,
\begin{equation}\label{2.6}
-\frac{1}{2}\bar{q}(\xi^{\dag}\td{\chi}\xi^{\dag}
 +\xi\td{\chi}^{\dag}\xi)q 
-\frac{\kappa}{2}\bar{q}(\xi^{\dag}\td{\chi}\xi^{\dag}
 -\xi\td{\chi}^{\dag}\xi)\gamma_5q,
\end{equation}
where $\td{\chi}=s+ip$, $s=s_{\rm ext}+{\cal M}$, ${\cal M}={\rm
diag}\{m_u,m_d,m_s\}$ is light current quark mass matrix, $s_{\rm ext}$
and $p$ are scalar and pseudoscalar external field respectively.
Eq.~(\ref{2.6}) will return to standard quark mass term of QCD,
$\bar{\psi}{\cal M}\psi$, in absence of pseudoscalar mesons at high
energy for arbitrary $\kappa$. Therefore, although the light current quark
masses are defined uniquely in this formalism, the symmetry and the
constrains of underlying QCD still can not fixed the coupling between
pseudoscalar mesons and constituent quarks. $\kappa$ will be treated as
an initial parameter of the model and be fitted phenomenologically. 

To conclude this section, the ChCQM lagrangian with light current quark
masses is
\begin{eqnarray}\label{2.7}
{\cal L}_{\chi}&=&i\bar{q}(\sla{\pa}+\sla{\Gamma}+
  g_{_A}{\slash\!\!\!\!\Delta}\gamma_5)q-m\bar{q}q
  -\frac{1}{2}\bar{q}(\xi^{\dag}\td{\chi}\xi^{\dag}
  +\xi\td{\chi}^{\dag}\xi)q
  -\frac{\kappa}{2}\bar{q}(\xi^{\dag}\td{\chi}\xi^{\dag}
 -\xi\td{\chi}^{\dag}\xi)\gamma_5q  \nonumber \\
   &&+\frac{F^2}{16}<\nabla_\mu U\nabla^\mu U^{\dag}>,
\end{eqnarray}
where vector meson fields have been omitted. The effects of isospin
breaking due to inequality of light quark masses will be generated via
constituent quark loops. For example, it generated masses of
pseudoscalar mesons and splits their decay constants.

\section{Quark Loops}

In this section, we will calculate pseudoscalar meson masses and decay
constants induced by one-loop effects of constituent quarks. They are the
leading order in $N_c^{-1}$ expansion.

In this framework, the effective action describing meson interaction can
be obtained via integrating over degrees of freedom of fermions
\begin{equation}\label{3.1}
e^{iS_{\rm eff}}\equiv\int{\cal D}\bar{q}{\cal D}qe^{i\int d^4x{\cal
   L}_\chi(x)}=<vac,out|in,vac>_{\rm Ext},
\end{equation}
where $<vac,out|in,vac>_{\rm Ext}$ is vacuum expectation value
in presence external sources. The above path integral can be performed
formally, and many methods, such as heat kernel
manner\cite{Sch51,Ball89}, have been used to regulate the bilinear
operator yielded via path integral. By using those formal integral
methods, however, the explicit calculations of high order contributions to
the chiral expansion are extremely tedious. In ref.\cite{Rho} we have
provided a convenient method to evaluate effective action via calculating
one-loop diagrams of constituent quarks directly. This method can capture
all high order contributions of the chiral expansion.

In interaction picture, the equation(~\ref{3.1}) is rewritten as follow
\begin{eqnarray}\label{3.2}
e^{iS_{\rm eff}}&=&<0|{\cal T}_qe^{i\int d^4x{\cal L}^{\rm I}_\chi(x)}|0>
       \nonumber \\ 
 &=&\sum_{n=1}^\infty i\int d^4p_1\frac{d^4p_2}{(2\pi)^4}
  \cdots\frac{d^4p_n}{(2\pi)^4}\tilde{\Pi}_n(p_1,\cdots,p_n)
  \delta^4(p_1-p_2-\cdots-p_n) \nonumber \\
&\equiv&i\Pi_1(0)+\sum_{n=2}^\infty i\int \frac{d^4p_1}{(2\pi)^4}
  \cdots\frac{d^4p_{n-1}}{(2\pi)^4}\Pi_n(p_1,\cdots,p_{n-1}),
\end{eqnarray}
where ${\cal T}_q$ is time-order product of constituent quark fields,
${\cal L}_{\chi}^{\rm I}$ is quark-meson interaction part of
lagrangian(~\ref{2.7}), $\tilde{\Pi}_n(p_1,\cdots,p_n)$ is one-loop
effects of constituent quarks with $n$ external sources,
$p_1,p_2,\cdots,p_n$ are four-momentas of $n$ external sources
respectively and
\begin{equation}\label{3.3}
\Pi_n(p_1,\cdots,p_{n-1})=\int d^4p_n\tilde{\Pi}_n(p_1,\cdots,p_n)
  \delta^4(p_1-p_2-\cdots-p_n).
\end{equation}
To get rid of all disconnected diagrams, we have
\begin{eqnarray}\label{3.4}
S_{\rm eff}&=&\sum_{n=1}^\infty S_n, \nonumber \\
S_1&=&\Pi_1(0),  \\
S_n&=&\int \frac{d^4p_1}{(2\pi)^4}\cdots\frac{d^4p_{n-1}}
  {(2\pi)^4}\Pi_n(p_1,\cdots,p_{n-1}), \hspace{0.8in}(n\geq 2)\nonumber.
\end{eqnarray}
Hereafter we will call $S_n$ as $n$-point effective action.

For the purpose of this paper, the lagrangian(~\ref{2.7}) can be
rewritten as follow
\begin{eqnarray}\label{3.5}
{\cal L}_{\chi}(x)={\cal L}_q(x)+{\cal L}_2^{(0)}(x)
  +{\cal L}[\pi(x),\eta_8(x)]+{\cal L}[K^\pm(x)]+{\cal L}[K^0(x)],
\end{eqnarray}
where ${\cal L}_q$ and ${\cal L}_2^{(0)}$ are free field lagrangian of
constituent quarks of pseudoscalar meson respectively,
\begin{eqnarray}\label{3.6}
{\cal L}_q&=&\sum_{i=u,d,s}\bar{q}_i(i\sla{\pa}-\bar{m}_i)q_i,
  \hspace{1in} \bar{m}_i=m+m_i, \nonumber \\
{\cal L}_2^{(0)}&=&\frac{F^2}{16}<\nabla_\mu U\nabla^\mu U^{\dag}>
               \nonumber \\
&=&\frac{F^2}{8}{\pa_\mu\Phi^a\pa^\mu\Phi^a+4a_\mu^a\pa^\mu\Phi^a+\cdots},  
  \hspace{0.5in}a=1,2,\cdots,8,     
\end{eqnarray}
where
\begin{eqnarray*}
   \lambda^a\Phi^a(x)&=&\sqrt{2}\left(\begin{array}{ccc}
    \frac{\pi_3}{\sqrt{2}}+\frac{\eta_8}{\sqrt{6}}&\pi^+&K^+   \\   
    \pi^-&-\frac{\pi_3}{\sqrt{2}}+\frac{\eta_8}{\sqrt{6}}&K^0   \\
    K^-&\bar{K}^0&-\frac{2}{\sqrt{3}}\eta_8
       \end{array} \right), \\
  \lambda^a a^a_\mu &=&\sqrt{2}\left(\begin{array}{ccc}
       A_\mu^{(u)}&a^+_\mu &A^{+}_\mu   \\   
    a^-_\mu& A_\mu^{(d)}&A^{0}_\mu   \\
       A^{-}_\mu&\bar{A}^{0}_\mu& A_\mu^{(s)}
       \end{array} \right).
\end{eqnarray*}
${\cal L}[\Phi(x)]$ denotes quark-meson coupling lagrangian,
\begin{eqnarray}\label{3.7}
{\cal L}[\pi,\eta_8]&=&\frac{g_A}{\sqrt{2}}[(\pa_\mu\pi^++2a_\mu^+)
   \bar{u}\gamma^\mu\gamma_5d+c.c.]+\frac{g_A}{2}\sum_{i=u,d,s}
   (\pa_\mu P_i+2A_{\mu}^{(i)})\bar{q}_i\gamma^\mu\gamma_5q_i 
     \nonumber \\
   &&+\frac{i}{\sqrt{2}}\kappa(m_u+m_d)(\pi^+\bar{u}\gamma_5d+c.c.)
  +i\kappa\sum_{i=u,d,s}m_iP_i\bar{q}_i\gamma_5q_i \nonumber \\
   &&+\frac{1}{2}(m_u+m_d)\pi^+\pi^-(\bar{u}u+\bar{d}d)
     +\frac{1}{2}\sum_{i=u,d,s}m_iP_i^2\bar{q}_iq_i,\\
{\cal L}[K^\pm]&=&\frac{g_A}{\sqrt{2}}[(\pa_\mu K^++2A_\mu^+)
   \bar{u}\gamma^\mu\gamma_5s+c.c.]
   +\frac{i}{\sqrt{2}}\kappa(m_u+m_s)(K^+\bar{u}\gamma_5s+c.c.)
      \nonumber \\
  &&+\frac{1}{2}(m_u+m_s)K^+K^-(\bar{u}u+\bar{s}s),\nonumber \\
{\cal L}[K^0]&=&\frac{g_A}{\sqrt{2}}[(\pa_\mu K^0+2A_\mu^0)
   \bar{d}\gamma^\mu\gamma_5s+(\pa_\mu\bar{K}^0+2\bar{A}_\mu^0)
   d\gamma^\mu\gamma_5\bar{s}]        \nonumber \\
   &&+\frac{i}{\sqrt{2}}\kappa(m_d+m_s)(K^0\bar{d}\gamma_5s
  +\bar{K}^0d\gamma_5\bar{s}) 
  +\frac{1}{2}(m_d+m_s)K^0\bar{K}^0(\bar{d}d+\bar{s}s),\nonumber      
\end{eqnarray}
where $c.c.$ denotes charge conjugate term of pervious term,  
\begin{eqnarray}\label{3.8}
P_u=\pi_3+\frac{1}{\sqrt{3}}\eta_8,\;\;\;\;\;\;\;\;
P_d=-\pi_3+\frac{1}{\sqrt{3}}\eta_8,\;\;\;\;\;\;\;\;
P_s=-\frac{2}{\sqrt{3}}\eta_8,
\end{eqnarray}
and $A_{\mu}^{(u)},\;A_{\mu}^{(d)};A_{\mu}^{(s)}$ etc. are axial-vector
external fields
correponding these pseudoscalar meson fields. From eq.(~\ref{3.7}) we can
see that the $K^\pm$-quark coupling and $K^0$-quark coupling are similar
to $\pi^\pm$-quark coupling. Thus we only need to calculate masses and
decay constants of $\pi^\pm,\;\pi^0$ and $\eta^8$. Then masses
and decay constants of $K^\pm$ can be obtained via replacing $m_d$ by
$m_s$ in one of $\pi^\pm$, and masses and decay constants of $K^0$ can
be obtained via replacing $m_u$ by $m_s$ in one of $\pi^\pm$.

The one-point effective action is generated by tadpole-loop of
constituent quarks. Calculation about this tadpole-loop contribution is
simple.
\begin{eqnarray}\label{3.9}
iS_1[\pi,\eta_8]&=&\frac{i}{2}\int d^4x\{(m_u+m_d)\pi^+(x)\pi^-(x)
  <0|T(\bar{u}(x)u(x)+\bar{d}(x)d(x))|0> \nonumber \\
  &&+\sum_{i=u,d,s}m_iP_i^2(x)<0|T\bar{q}_i(x)q_i(x)|0>\nonumber \\
&=&-\frac{i}{2}\int d^4x\int\frac{d^4k}{(2\pi)^4}\{
  [(m_u+m_d)\pi^+(x)\pi^-(x)+m_uP_1^2(x)]Tr[S_F^{(u)}(k)]
   \nonumber \\
  &&+[(m_u+m_d)\pi^+(x)\pi^-(x)+m_dP_2^2(x)]Tr[S_F^{(d)}(k)]
  +m_sP_3^2(x)Tr[S_F^{(s)}(k)]\},
\end{eqnarray}
where $Tr$ denotes trace taking over color and Lorentz space, $S_F^{(q)}$
is propagator of constituent quark fields, 
$S_F^{(q)}(k)=i(\sla{k}-\bar{m}_q+i\ep)^{-1}$. In terms of dimensional
regularization, we can integrate over internal line momenta $k$ in the
above equation. The result is
\begin{eqnarray}\label{3.10}
iS_1[\pi,\eta_8]&=&-\frac{2N_c}{(4\pi)^{D/2}}\Gamma(1-\frac{D}{2})
 i\int d^4x\{(\frac{\mu^2}{\bar{m}_u^2})^{\ep/2}\bar{m}_u^3
  [(m_u+m_d)\pi^+\pi^-+m_uP_1^2] \nonumber \\
&&+(\frac{\mu^2}{\bar{m}_d^2})^{\ep/2}\bar{m}_d^3
   [(m_u+m_d)\pi^+\pi^-+m_dP_2^2]
 +(\frac{\mu^2}{\bar{m}_s^2})^{\ep/2}\bar{m}_s^3m_sP_3^2\}.
\end{eqnarray}
Defining a constant $B_0$ to absorbe the quadratic divergence from loop
integral,
\begin{equation}\label{3.11}
\frac{F_0^2}{16}B_0=\frac{N_c}{(4\pi)^{D/2}}(\frac{\mu^2}{m^2})^{\ep/2}
    \Gamma(1-\frac{D}{2})m^3,
\end{equation}
we have
\begin{eqnarray}\label{3.12}
S_1[\pi,\eta_8]&=&\int d^4x{\cal L}_1[\pi(x),\eta_8(x)]\nonumber \\
{\cal L}_1[\pi,\eta_8]&=&-\frac{F_0^2}{8}B_0\{(x_u^3+x_d^3)(m_u+m_d)
  \pi^+\pi^-+\sum_{i=u,d,s}x_i^3m_iP_i^2\} \nonumber \\
&&-\frac{N_c}{8\pi^2}m^3\{(x_u^3\ln{x_u^2}+x_d^3\ln{x_d^2})(m_u+m_d)
  \pi^+\pi^-+\sum_{i=u,d,s}m_ix_i^3\ln{x_i^2}P_i^2\}
\end{eqnarray}
where $x_q=\bar{m}_q/m$.

The two-point effective action concerning to masses and decay constants
of charge pion can be obtained as follow,
\begin{eqnarray}\label{3.13}
&&iS_2[\pi^\pm]
=\frac{g_A^2}{2}\int\frac{d^4q}{(2\pi)^4}\frac{d^4k}{(2\pi)^4}
  (iq_\mu\pi^+(q)+2a_\mu^+(q))(-iq_\nu\pi^-(-q)+2a_\nu^-(-q))
  Tr[\gamma^\mu\gamma_5S_F^{(d)}(k-q)\gamma^\nu\gamma_5S_F^{(u)}(k)]
     \nonumber \\
&&\;\;\;\;\;\;+\frac{i}{2}g_A\kappa(m_u+m_d)
  \int\frac{d^4q}{(2\pi)^4}\frac{d^4k}{(2\pi)^4}\{
  (iq_\mu\pi^+(q)+2a_\mu^+(q))\pi^-(-q)
  Tr[\gamma^\mu\gamma_5S_F^{(d)}(k-q)\gamma_5S_F^{(u)}(k)]+c.c.\}
   \nonumber \\
&&\;\;\;\;\;\;-\frac{\kappa^2}{2}(m_u+m_d)^2\int\frac{d^4q}{(2\pi)^4}
 \frac{d^4k}{(2\pi)^4}\pi^+(q)\pi^-(-q)
 Tr[\gamma_5S_F^{(d)}(k-q)\gamma_5S_F^{(u)}(k)]\nonumber \\
&=&\frac{2N_c}{(4\pi)^2}\Gamma(2-\frac{D}{2})g_A^2
 i\int d^4x\int\frac{d^4q}{(2\pi)^4}
e^{iq\cdot x}(iq_\mu\pi^+(q)+2a_\mu^+(q))(\pa^\mu\pi^-(x)+2a^{-\mu}(x))
   \nonumber \\&&\;\;\;\;\;\;\;\;\;\;\times
\int_0^1dt[\bar{m_u}^2+\bar{m}_u\bar{m}_d+t(\bar{m}_d^2-\bar{m}_u^2)]
\left(\frac{\bar{m_u}^2+t(\bar{m}_d^2-\bar{m}_u^2)-t(1-t)q^2}
 {4\pi\mu^2}\right)^{-\frac{\ep}{2}}\nonumber \\
&&+\frac{2N_c}{(4\pi)^2}\Gamma(2-\frac{D}{2})
 \kappa g_A(m_u+m_d)i\int d^4x\int\frac{d^4q}{(2\pi)^4}
  e^{iq\cdot x}[(iq_\mu\pi^+(q)+2a_\mu^+(q))\pa^\mu\pi^-(x)+c.c.]
     \nonumber \\&&\;\;\;\;\;\;\;\;\;\;\times
\int_0^1dt[\bar{m_u}+t(m_d-m_u)]
\left(\frac{\bar{m_u}^2+t(\bar{m}_d^2-\bar{m}_u^2)-t(1-t)q^2}
 {4\pi\mu^2}\right)^{-\frac{\ep}{2}} \\
&&+\frac{2N_c}{(4\pi)^2}\kappa^2(m_u+m_d)^2
i\int d^4x\int\frac{d^4q}{(2\pi)^4}
 e^{iq\cdot x}\pi^+(q)\pi^-(x)\int_0^1dt
\left(\frac{\bar{m_u}^2+t(\bar{m}_d^2-\bar{m}_u^2)-t(1-t)q^2}
 {4\pi\mu^2}\right)^{-\frac{\ep}{2}}
      \nonumber\\&&\;\;\;\;\;\;\;\;\;\;\times
\{\Gamma(1-\frac{D}{2})[\bar{m_u}^2+t(\bar{m}_d^2-\bar{m}_u^2)
  -t(1-t)q^2]
-\Gamma(2-\frac{D}{2})[\bar{m_u}^2-\bar{m}_u\bar{m}_d
  +t(\bar{m}_d^2-\bar{m}_u^2)-2t(1-t)q^2]\}. \nonumber 
\end{eqnarray}
There are both quadratic divergence and logarithmic divergence in the
above effective action. The quadratic divergence can be canceled by
constant $B_0$ defined in eq.(~\ref{3.11}), and the logarithmic divergence
can be conceled via defining
\begin{eqnarray}\label{3.14}
g^2&=&\frac{8N_c}{3(4\pi)^2}\Gamma(2-\frac{D}{2})(\frac{4\pi\mu^2}{m^2})
  ^{\frac{\ep}{2}},  \nonumber \\
\frac{F_0^2}{16}&=&\frac{F^2}{16}+\frac{N_c}{(4\pi)^2}g_A^2m^2
  \Gamma(2-\frac{D}{2})(\frac{4\pi\mu^2}{m^2})^{\frac{\ep}{2}}.
\end{eqnarray}
$g$ is an universal coupling constant of this model. In ref.\cite{Rho}, it
has been determined as $g^2=\frac{N_c}{3\pi^2}$ by the first KSRF sum
rule\cite{KSRF}.

Then eq.(~\ref{3.13}) together with eq.(~\ref{3.12}) give $O(N_c)$
effective lagrangian containing the terms linear or quadratic in the
charge pion fields as follow	
\begin{eqnarray}\label{3.15}
{\cal L}_2[\pi^\pm(x)]&=&\frac{F^2(m_u,m_d)}{4}\pa_\mu\pi^+\pa^\mu\pi^-
  +\frac{\bar{f}^2(m_u,m_d)}{2}(a_\mu^+\pa^\mu\pi^-+c.c.)
  -\frac{F_0^2}{4}\bar{M}^2(m_u,m_d)\pi^+\pi^-\nonumber \\
 &&+\int\frac{d^4q}{(2\pi)^4}
 e^{iq\cdot x}\{\frac{\alpha(q^2;m_u,m_d)}{2}iq^\mu(a_\mu^+(x)\pi^-(q)
  +c.c.)-\frac{F_0^2}{4}\beta(q^2;m_u,m_d))\pi^+(q)\pi^-(x)\},
\end{eqnarray}
where
\begin{eqnarray}\label{3.16}
&&F^2(m_u,m_d)=F_0^2+\frac{3}{2}g^2g_A^2(m_u+m_d)(4m+m_u+m_d)
  +3g^2\kappa g_A(m_u+m_d)(\bar{m}_u+\bar{m}_d)\nonumber \\
&&\;\;\;\;\;\;\;\;-\frac{N_c}{2\pi^2}g_A
 [g_A(\bar{m}_u+\bar{m}_d)+2\kappa(m_u+m_d)]
\int_0^1dt[\bar{m}_u+t(m_d-m_u)]
\ln{(x_u^2+t(x_d^2-x_u^2))}\nonumber\\
&&\;\;\;\;\;\;\;\;
+4\kappa^2(m_u+m_d)^2\{(\frac{g^2}{4}-\frac{F_0^2B_0}{48m^3}
  -\frac{N_c}{24\pi^2})-\frac{N_c}{8\pi^2}\int_0^1dt\cdot t(1-t)
[3\ln{(x_u^2+t(x_d^2-x_u^2))}
-\frac{\bar{m}_u\bar{m}_d}{\bar{m}_u^2+t(\bar{m}_d^2-\bar{m}_u^2)}]\},
   \nonumber\\
&&\bar{f}^2(m_u,m_d)=F_0^2+\frac{3}{2}g^2g_A^2(m_u+m_d)(4m+m_u+m_d)
  +\frac{3}{2}g^2\kappa g_A(m_u+m_d)(\bar{m}_u+\bar{m}_d)\nonumber \\
&&\;\;\;\;\;\;\;\;-\frac{N_c}{2\pi^2}g_A
 [g_A(\bar{m}_u+\bar{m}_d)+\kappa(m_u+m_d)]
\int_0^1dt[\bar{m}_u+t(m_d-m_u)]
\ln{(x_u^2+t(x_d^2-x_u^2))},\nonumber\\
&&\bar{M}^2(m_u,m_d)=B_0(m_u+m_d)\{\frac{1}{2}(x_u^3+x_d^3)
 +\frac{N_c}{2\pi^2}\frac{m^3}{F_0^2B_0}(x_u^3\ln{x_u^2}+x_d^3\ln{x_d^3})
 -\frac{\kappa^2}{4}\frac{m_u+m_d}{m}(x_u^2+x_d^2)\}\nonumber \\
&&\;\;\;\;\;\;\;\;+\frac{3}{2}g^2\kappa^2\frac{(m_d^2-m_u^2)^2}{F_0^2}
 -\frac{N_c}{2\pi^2}\kappa^2\frac{(m_u+m_d)^2}{F_0^2}\int_0^1dt
 [2\bar{m}_u^2-\bar{m}_u\bar{m}_d+2t(\bar{m}_d^2-\bar{m}_u^2)]
 \ln{(x_u^2+t(x_d^2-x_u^2))},\\
&&\alpha(q^2;m_u,m_d)=-\frac{N_c}{2\pi^2}g_A
 [g_A(\bar{m}_u+\bar{m}_d)+\kappa(m_u+m_d)]
\int_0^1dt[\bar{m}_u+t(m_d-m_u)]
\ln{\left(1-\frac{t(1-t)q^2}{\bar{m}_u^2+t(\bar{m}_d^2-\bar{m}_u^2)}
 \right)},\nonumber \\
&&\beta(q^2;m_u,m_d)=\frac{N_c}{2\pi^2}g_A\frac{q^2}{F_0^2}
 [g_A(\bar{m}_u+\bar{m}_d)+2\kappa(m_u+m_d)]
\int_0^1dt[\bar{m}_u+t(m_d-m_u)]
\ln{\left(1-\frac{t(1-t)q^2}{\bar{m}_u^2+t(\bar{m}_d^2-\bar{m}_u^2)}
 \right)}\nonumber \\&&\;\;\;\;\;\;\;\;
-\frac{N_c}{2\pi^2}\kappa^2\frac{(m_u+m_d)^2}{F_0^2}\int_0^1dt
\{[2\bar{m}_u^2-\bar{m}_u\bar{m}_d+2t(\bar{m}_d^2-\bar{m}_u^2)
  -3t(1-t)q^2]\ln{\left(1-\frac{t(1-t)q^2}{\bar{m}_u^2
  +t(\bar{m}_d^2-\bar{m}_u^2)}\right)}\nonumber \\&&\hspace{2in}
  +t(1-t)q^2\left(2-\frac{\bar{m}_u\bar{m}_d}{\bar{m}_u^2
  +t(\bar{m}_d^2-\bar{m}_u^2)}\right)\}. \nonumber 
\end{eqnarray}
It should be pointed out that $\alpha(q^2;m_u,m_d)$ is order $q^2$ at
least and $\beta(q^2;m_u,m_d)$ is order $q^4$ at least. Since in this
paper we focus on pseudoscalar meson spectrums, these high order
derivative terms
should obey motion equation of pseudoscalar mesons. In momentum space, the
motion equation of physical pseudoscalar mesons is generally written
\begin{eqnarray}\label{3.17}
(q^2-m_{\vphi}^2)\vphi(q)=-if_{\vphi}q^\mu A_\mu^{(\vphi)}(-q),
\end{eqnarray}
where $m_{\vphi}$ and $f_{\vphi}$ are physical mass and decay constants of
pseudoscalar, e.g., $m_\pi=135$MeV and $f_\pi=185.2$MeV. Due to this
motion equation, we have 
\begin{eqnarray}\label{3.18}
\alpha(q^2;m_u,m_d)a_\mu^+(x)\pi^-(q)
&=&\alpha(m_{\pi}^2;m_u,m_d)a_\mu^+(x)\pi^-(q),\nonumber \\
\beta(q^2;m_u,m_d)\pi^+(q)\pi^-(-q)
&=&\beta(m_\pi^2;m_u,m_d)\pi^+(q)\pi^-(-q)-\frac{i}{2}q^\mu f_\pi
  \beta'(m_{\pi}^2;m_u,m_d)(a_\mu^+(x)\pi^-(q)+c.c.),
\end{eqnarray}
where
\begin{eqnarray}\label{3.19}
\beta'(m_{\pi}^2;m_u,m_d)=\frac{d}{dq^2}
 \beta'(q^2;m_u,m_d)|_{q^2=m_{\pi}^2}.
\end{eqnarray}
Thus eq.(~\ref{3.15}) can be rewritten
\begin{eqnarray}\label{3.20}
{\cal L}_2[\pi^\pm(x)]&=&\frac{F^2(m_u,m_d)}{4}\pa_\mu\pi^+\pa^\mu\pi^-
-\frac{F_0^2}{4}\{\bar{M}^2(m_u,m_d)+\beta(m_{\pi}^2;m_u,m_d)\}\pi^+\pi^-
   \nonumber \\
&&+\{\frac{\bar{f}^2(m_u,m_d)}{2}+\frac{\alpha(m_\pi^2;m_u,m_d)}{2}
   +\frac{F_0^2}{4F(m_u,m_d)}f_\pi\beta'(m_{\pi}^2;m_u,m_d)\}
  (a_\mu^+\pa^\mu\pi^-+c.c.).
\end{eqnarray}

The two-point effective action concerning to masses and decay constants of
neutral pion and $\eta_8$ can be evaluated similarly. However the decay
constants for neutral mesons cannot be extracted directly from the data.
It means that the decay constants for neutral mesons cannot be used
to determine light current quark masses. Therefore, in this paper we do
not need to evaluate the decay constants for neutral pion and $\eta_8$.
The effective action containing the quadratic terms of neutral pion and
$\eta_8$ is then
\begin{eqnarray}\label{3.21}
&&iS_2[\pi^0,\eta_8] \nonumber \\
&=&\frac{g_A^2}{8}\int\frac{d^4q}{(2\pi)^4}
\frac{d^4k}{(2\pi)^4}\sum_{j=u,d,s}q^2 P_j(q)P_j(-q)
 Tr[\gamma^\mu\gamma_5S_F^{(j)}(k-q)\gamma^\nu\gamma_5S_F^{(j)}(k)] 
  \nonumber \\
&&+\frac{i}{2}g_A\kappa\sum_{j=u,d,s}m_j\int\frac{d^4q}{(2\pi)^4}
\frac{d^4k}{(2\pi)^4}(iq_\mu)P_j(q)P_j(-q)
 Tr[\gamma^\mu\gamma_5S_F^{(j)}(k-q)\gamma_5S_F^{(j)}(k)] \nonumber \\
&&-\frac{\kappa^2}{2}\sum_{j=u,d,s}m_j^2\int\frac{d^4q}{(2\pi)^4}
  \frac{d^4k}{(2\pi)^4}P_j(q)P_j(-q)
 Tr[\gamma_5S_F^{(j)}(k-q)\gamma_5S_F^{(j)}(k)] \nonumber \\
&=&\frac{N_c}{(4\pi)^2}\Gamma(2-\frac{D}{2})g_A^2i\int
 \frac{d^4q}{(2\pi)^4}q^2\sum_{j=u,d,s}\bar{m}_j^2
 q^2 P_j(q)P_j(-q)\int_0^1dt\left(\frac{\mu^2}{\bar{m}_u^2-t(1-t)q^2}
 \right)^{\frac{\ep}{2}}\nonumber \\
&&+\frac{2N_c}{(4\pi)^2}\Gamma(2-\frac{D}{2})\kappa g_Ai
 \int\frac{d^4q}{(2\pi)^4}q^2\sum_{j=u,d,s}m_j\bar{m}_j
 q^2 P_j(q)P_j(-q)\int_0^1dt\left(\frac{\mu^2}{\bar{m}_u^2-t(1-t)q^2}
 \right)^{\frac{\ep}{2}}\nonumber \\
&&+\frac{2N_c}{(4\pi)^2}\kappa^2 i\int\frac{d^4q}{(2\pi)^4}\sum_{j=u,d,s}
  m_j^2P_j(q)P_j(-q)\int_0^1dt\left(\frac{\mu^2}{\bar{m}_u^2-t(1-t)q^2}
  \right)^{\frac{\ep}{2}}\nonumber \\&&\hspace{1in}\times
\{\Gamma(1-\frac{D}{2})[\bar{m}_j^2-t(1-t)q^2]+2t(1-t)q^2
 \Gamma(2-\frac{D}{2})\}.
\end{eqnarray}
The divergences in the above effective action can be canceled by
eqs.(~\ref{3.11}) and (~\ref{3.14}). Then from eqs.(~\ref{3.12}) and
(~\ref{3.21}) we can obtain $O(N_c)$ effective lagrangian describing
two-point vertex of $\pi^0$ and $\eta_8$ as follow
\begin{eqnarray}\label{3.22}
{\cal L}_2[\pi^0(x),\eta_8(x)]=\sum_{i=u,d,s}\{\frac{F_i^2(m_i)}{8}
  \pa_\mu P_i\pa^\mu P_i-\frac{F_0^2}{8}\bar{M}_i^2(m_i)P_i^2
  -\frac{F_0^2}{8}\int\frac{d^4q}{(2\pi)^4}e^{iq\cdot x}
  \beta_i(q^2;m_i)P_i(q)P_i(x)\},
\end{eqnarray}
where
\begin{eqnarray}\label{3.23}
F_i^2(m_i)&=&\frac{F_0^2}{2}+3g^2g_A^2m_i(2m+m_i)+6g^2\kappa g_Am_i
  \bar{m}_i+8\kappa^2(\frac{g^2}{4}-\frac{F_0^2B_0}{48m^3}
  -\frac{N_c}{48\pi^2})m_i^2\nonumber \\
&&-\frac{N_c}{2\pi^2}(g_A^2\bar{m}_i^2+2g_A\kappa m_i\bar{m}_i
   +\kappa^2 m_i^2)\ln{x_i^2}, \nonumber \\
\bar{M}_i^2(m_i)&=&(B_0+\frac{N_c}{\pi^2}\frac{m^3}{F_0^2}\ln{x_i^2})
    m_ix_i^2(x_i-\frac{m_i}{m}\kappa^2), \\
\beta_i(q^2;m_i)&=&\frac{N_c}{2\pi^2F_0^2}q^2\{\frac{\kappa^2}{3}m_i^2
 +\left(g_A^2\bar{m}_i^2+2\kappa m_i\bar{m}_i+2\kappa^2 m_i^2
  [\bar{m}_i^2q^{-2}-3t(1-t)]\right)\int_0^1dt
  \ln{(1-\frac{t(1-t)q^2}{\bar{m}_i^2})}\}.  \nonumber 
\end{eqnarray}
Since auxiliary fields $P_i$ do not lie in physical hadron spectrums, the
equation of motion (\ref{3.17}) can not be used in (\ref{3.22}) simply. In
section 5, we will use propagator method to deal with the terms with high
power momenta in (\ref{3.22}) and diagonalize $\pi_3-\eta_8$ mixing. 

\section{Meson Loops and Renormaliztion}

Purpose of this section is to evaluate one-loop effects of pseudoscalar
mesons. Due to parity conservation, there are only tadpole diagrams of
pseudoscalar mesons contributing to masses of decay constants of $0^-$
mesons(fig. 1). Since ChCQM is a nonrenormalizable effective theory, it is
very difficlut to calculate meson loop effects completely. However, we can
expect that, in mass spectrums and decay constants of pseudoscalar mesons, 
the dominant one-loop effects is generated by the lowest order effective
lagrangian, because neglect is $O(m_q^3)$ and suppressed by $N_c^{-1}$
expansion. Fortunately, the formalism is renormilzed up to this
order. 

The lowest order effective lagrangian is well-known
\begin{eqnarray}\label{4.1}
{\cal L}_2&=&\frac{F_0^2}{16}<\nabla_\mu U\nabla^\mu U^{\dag}
  +\chi U^{\dag}+U\chi^{\dag}> \nonumber \\
  &=&\frac{F_0^2}{48}<[\lambda^a,\Delta_\mu][\lambda^a,\Delta^\mu]>
    +\frac{3F_0^2}{256}<\lambda^a\lambda^a(\xi\chi^{\dag}\xi
     +\xi^{\dag}\chi\xi^{\dag})>,
\end{eqnarray}
where $\lambda^a(a=1,2,\cdots,8)$ are Gell-Mann matrices, 
$\chi=2B_0\td{\chi}$ and the following SU(N) completeness relation have
been used
\begin{eqnarray}\label{4.2}
\sum_{a=1}^{N^2-1}<\lambda^aA\lambda^aB>&=&-\frac{2}{N}<AB>+2<A><B>,
  \nonumber \\
\sum_{a=1}^{N^2-1}<\lambda^aA><\lambda^aB>&=&\;   
  2<AB>-\frac{2}{N}<A><B>.
\end{eqnarray}
To evaluate the one-loop graphs generated by this lagrangian, we consider
the quantum fluctuation $\vphi(x)=\vphi^a(x)\lambda^a$ around the
solution $\bar{U}(x)=\bar{\xi}^2(x)$ to the classical equations of motion,
\begin{eqnarray}\label{4.3}
U=\bar{\xi}e^{i\vphi}\bar{\xi}.
\end{eqnarray}

Substituting expansion(~\ref{4.3}) into ${\cal L}_2$ and retaining terms
up to and including $\vphi^2$ one obtains
\begin{eqnarray}\label{4.4}
{\cal L}_2\rightarrow \bar{\cal L}_2+\frac{F_0^2}{8}(\pa_\mu\vphi^a
 \pa^\mu\vphi^a-m_\vphi^2\vphi^a\vphi^a)-\frac{F_0^2}{16}
 <[\vphi,\Delta_\mu][\vphi,\Delta^\mu]+\frac{1}{4}\{\vphi,\vphi\}
 (\bar{\xi}\chi^{\dag}\bar{\xi}+\bar{\xi}^{\dag}\chi\bar{\xi}^{\dag}
 -2{\cal M})>,
\end{eqnarray}
where we have omitted some terms which do not contribute to masses and
decay constants via one-loop graphs. The contribution of tadpole graphs
can be calculated easily
\begin{eqnarray}\label{4.5}
{\cal L}_2^{(\rm tad)}&=&-\frac{1}{4}
 <[\lambda^a,\Delta_\mu][\lambda^a,\Delta^\mu]+\frac{1}{4}
 \{\lambda^a,\lambda^a\}(\bar{\xi}\chi^{\dag}\bar{\xi}
  +\bar{\xi}^{\dag}\chi\bar{\xi}^{\dag}-2{\cal M})>
 \int\frac{d^4k}{(2\pi)^4}\frac{i}{k^2-m_\vphi^2+i\ep} \nonumber \\
&=&\frac{1}{4}(m_\vphi^2\lambda-\frac{m_\vphi^2}{16\pi^2}
  \ln{\frac{m_\vphi^2}{\mu^2}})
  <[\lambda^a,\Delta_\mu][\lambda^a,\Delta^\mu]+\frac{1}{4}
 \{\lambda^a,\lambda^a\}(\bar{\xi}\chi^{\dag}\bar{\xi}
  +\bar{\xi}^{\dag}\chi\bar{\xi}^{\dag}-2{\cal M})>,
\end{eqnarray}
where 
$$\lambda=\frac{1}{16\pi^2}\{\frac{2}{\ep}+\ln{(4\pi)}+\gamma+1\}.$$
Comparing eq.(~\ref{4.5}) and eq.(~\ref{4.1}) we can see that the
divergence $\lambda$ can be absorbed by free parameters $F_0$ and $B_0$.
Thus the sum of tree graphs and tadpole contribution is
\begin{eqnarray}\label{4.6}
{\cal L}_2^{(t)}=\frac{F_0^2}{48}(1-3\mu_\vphi)
v <[\lambda^a,\Delta_\mu][\lambda^a,\Delta^\mu]>
 +\frac{3F_0^2}{256}(1-\frac{8}{3}\mu_\vphi)<\lambda^a\lambda^a
  (\xi\chi^{\dag}\xi+\xi^{\dag}\chi\xi^{\dag})>,
\end{eqnarray}
where 
\begin{eqnarray}\label{4.7}
\mu_\vphi=\frac{m_\vphi^2}{4\pi^2F_0^2}\ln{\frac{m_\vphi^2}{\mu^2}}.
\end{eqnarray}
In lagrangian(~\ref{4.6}) we appoint that $m_\vphi=m_\pi$ when $a=1,2,3$,
$m_\vphi=m_{_K}$ when $a=4,5,6,7$ and $m_\vphi=m_{\eta_8}$ when $a=8$.

\section{Light Quark Mass Determination Beyond the Chiral
Perturbation Expansion}

For extracting $(m_{_{K^+}})_{\rm QCD}$
from experimental data, the electromagnetic mass splitting of $K$-meson is
required. The prediction of Dashen theorem\cite{Dashen69},
$(m_{_{K^+}}-m_{_{K^0}})_{e.m.}=1.3$MeV, has been corrected in serveral
recent analysis with considering contribution from vector meson exchange.
A larger correction is first obtained by Donoghue, Holstein and
Wyler\cite{DHW93}, who find $(m_{_{K^+}}-m_{_{K^0}})_{e.m.}=2.3$MeV. Then
Bijnens and Prades\cite{Bijnens97}, who evaluated both long-distance
contribution using ENJL model and short-distance contribution using
perturbative QCD and factorization, find 
$(m_{_{K^+}}-m_{_{K^0}})_{e.m.}=2.4\pm 0.3$MeV at $\mu=m_\rho$. Gao
{\sl et.al.}\cite{Gao97} also gave
$(m_{_{K^+}}-m_{_{K^0}})_{e.m.}=2.5$MeV.
Baur and Urech\cite{BU96}, however, obtained a smaller correction, 
$(m_{_{K^+}}-m_{_{K^0}})_{e.m.}=1.6$MeV at $\mu=m_\rho$. In addition,
calculation of lattice QCD\cite{Duncan96} found
$(m_{_{K^+}}-m_{_{K^0}})_{e.m.}=1.9$MeV. 
These estimates indicate that the corrections to Dashen theorem are indeed
substantial. In this the present paper, we average the above results and
take $m_{_{K^+}}-m_{_{K^0}})_{e.m.}=2.1\pm 0.1$MeV at energy scale
$\mu=m_\rho$.

From eqs.(~\ref{3.20}) and (\ref{4.6}), the masses and decay constants of
koan and chrage pion can be obtained via solve the following equations
\begin{eqnarray}\label{5.1}
m_{\pi^+}^2&=&\frac{F_0^2}{F_{_R}^2(m_u,m_d)}\{\bar{M}_{_R}^2(m_u,m_d)
  +\beta(m_\pi^2;m_u,m_d)\}
   \nonumber \\
m_{_{K^+}}^2&=&\frac{F_0^2}{F_{_R}^2(m_u,m_s)}\{\bar{M}_{_R}^2(m_u,m_s) 
  +\beta(m_{_K}^2;m_u,m_s)\},
   \nonumber \\
m_{_{K^0}}^2&=&\frac{F_0^2}{F_{_R}^2(m_d,m_s)}\{\bar{M}_{_R}^2(m_d,m_s)
  +\beta(m_{_K}^2;m_d,m_s)\}, \\
f_{\pi^+}&=&\frac{\bar{f}_{_R}^2(m_u,m_d)}{F_{_R}(m_u,m_d)}+
  \frac{\alpha(m_\pi^2;m_u,m_d)}{F_{_R}(m_u,m_d)}
  +\frac{F_0^2}{2F_{_R}^2(m_u,m_d)}f_{\pi^+}\beta'(m_\pi^2;m_u,m_d),
   \nonumber \\
f_{_{K^+}}&=&\frac{\bar{f}_{_R}^2(m_u,m_s)}{F_{_R}(m_u,m_s)}+
  \frac{\alpha(m_{_K}^2;m_u,m_s)}{F_{_R}(m_u,m_s)}
  +\frac{F_0^2}{2F_{_R}^2(m_u,m_s)}f_{_K}\beta'(m_{_K}^2;m_u,m_s),
    \nonumber \\
f_{_{K^0}}&=&\frac{\bar{f}_{_R}^2(m_d,m_s)}{F_{_R}(m_d,m_s)}+
  \frac{\alpha(m_{_K}^2;m_d,m_s)}{F_{_R}(m_d,m_s)}
  +\frac{F_0^2}{2F_{_R}^2(m_d,m_s)}f_{_K}\beta'(m_{_K}^2;m_d,m_s),
  \nonumber
\end{eqnarray}
where subscript ``R'' denotes renormalized quantity,
\begin{eqnarray}\label{5.2}
F_{_R}^2(m_u,m_d)&=&F^2(m_u,m_d)-F_0^2(2\mu_\pi+\mu_{_K}), \nonumber \\
F_{_R}^2(m_i,m_s)&=&F^2(m_i,m_s)-\frac{3}{4}F_0^2(\mu_\pi+2\mu_{_K}
    +\mu_{\eta_8}), \hspace{1in} (i=u,d)\nonumber \\
\bar{M}_{_R}^2(m_u,m_d)&=&\bar{M}^2(m_u,m_d)
  -B_0(m_u+m_d)(\frac{3}{2}\mu_\pi
  +\mu_{_K}+\frac{1}{6}\mu_{\eta_8}), \nonumber \\
\bar{M}_{_R}^2(m_i,m_s)&=&\bar{M}^2(m_i,m_s)-B_0(m_i+m_s)
  (\frac{3}{4}\mu_\pi+\frac{3}{2}\mu_{_K}+\frac{5}{12}\mu_{\eta_8}), \\  
\bar{f}_{_R}^2(m_u,m_d)&=&\bar{f}^2(m_u,m_d)-F_0^2(2\mu_\pi+\mu_{_K}), 
  \nonumber \\
\bar{f}_{_R}^2(m_i,m_s)&=&\bar{f}^2(m_i,m_s)-
   \frac{3}{4}F_0^2(\mu_\pi+2\mu_{_K}+\mu_{\eta_8}), \nonumber
\end{eqnarray}
Here the quantity $\mu_\vphi$ is defined in (\ref{4.7}) and depend on the 
renormalization scale $\mu$. It has been recognized that the scale plays
an important role in low-energy QCD. In this formalism, many parameters,
such as light current quark masses, constituent quark mass $m$,
axial-vector coupling constant $g_A$, are scale-dependent. The
characteristic scale of the model is described by the universal coupling
constant $g$ which is determined by the first KSRF sum rule at energy
scale $\mu=m_\rho$. Hence we take $\mu=m_\rho=770$MeV in $\mu_\vphi$, and
the physical masses and decay constants, however, should be independent of
the renormalization scale.

In order to obtain the masses of $\pi^0$ and $\eta^8$, the $\pi_3-\eta_8$
mixing in eq.(~\ref{3.22}) must be diagonalized. Eq.(\ref{3.22})
together with eq.(\ref{4.6}) lead to the quadratic terms for the
$\pi_3$ and $\eta_8$ are of the form
\begin{eqnarray}\label{5.3}
S_2[\pi_3,\eta_8]=\int\frac{d^4q}{(2\pi)^4}\{
  \frac{1}{2}(q^2-M_3^2(q^2))\pi_3^2
  +\frac{1}{2}(q^2-M_8^2(q^2))\eta_8^2-M_{38}^2(q^2)\pi_3\eta_8\}
\end{eqnarray}
where
\begin{eqnarray}\label{5.4}
M_3^2(q^2)&=&\frac{F_0^2}{F_3^2}\{\bar{M}_u^2(m_u)+\bar{M}_d^2(m_d)
   +\beta_u(q^2;m_u)+\beta_d(q^2;m_d)-2B_0\hat{m}
   (\frac{3}{2}\mu_\pi+\mu_{_K}+\frac{1}{6}\mu_{\eta_8})\}, \nonumber \\
M_8^2(q^2)&=&\frac{F_0^2}{3F_8^2}\{\bar{M}_u^2(m_u)+
  \bar{M}_d^2(m_d)+4\bar{M}_s^2(m_s)+\beta_u(q^2;m_u)+\beta_d(q^2;m_d)
  +4\beta_s(q^2;m_s) \nonumber \\ &&-2B_0(\hat{m}+2m_s)
  (2\mu_{_K}+\frac{2}{3}\mu_{\eta_8})-2B_0\hat{m}(-\frac{3}{2}\mu_\pi
  +\mu_{_K}+\frac{1}{2}\mu_{\eta_8})\},  \\
M_{38}^2(q^2)&=&\frac{F_0^2}{\sqrt{3}F_3F_8}\{\bar{M}_u^2(m_u)
   -\bar{M}_d^2(m_d)+\beta_u(q^2;m_u)-\beta_d(q^2;m_d)-B_0(m_u-m_d)
   (\frac{3}{2}\mu_\pi+\mu_{_K}+\frac{1}{6}\mu_{\eta_8})\}\nonumber \\
  &&+q^2\frac{F_u^2(m_u)-F_d^2(m_d)}{\sqrt{3}F_3F_8},\nonumber 
\end{eqnarray}
with $\hat{m}=(m_u+m_d)/2$ and
\begin{eqnarray}\label{5.5}
F_3^2&=&F_u^2(m_u)+F_d^2(m_d)-F_0^2(2\mu_\pi+\mu_{_K}),\nonumber \\
F_8^2&=&\frac{1}{3}\{F_u^2(m_u)+F_d^2(m_d)+4F_s^2(m_s)\}
    -3F_0^2\mu_{_K}.
\end{eqnarray}

Due to $\pi_3-\eta_8$ mixing, the ``physical'' propagators of $\pi^0$ and
$\eta_8$ are obtained via the chain approximation in momentum space
\begin{eqnarray}\label{5.6}
\frac{i}{q^2-m_{\pi^0}^2+i\ep}&=&\frac{i}{q^2-M_3^2(q^2)+i\ep}
  +\frac{iM_{38}^4(q^2)}{(q^2-M_8^2(q^2)+i\ep)(q^2-M_3^2(q^2)+i\ep)^2}
  +\cdots \nonumber \\
 &=&\frac{i}{q^2-M_3^2(q^2)-\frac{M_{38}^4(q^2)}{q^2-M_8^2(q^2)}+i\ep},
     \nonumber \\
\frac{i}{q^2-m_{\eta_8}^2+i\ep}&=&\frac{i}{q^2-M_8^2(q^2)+i\ep}
  +\frac{iM_{38}^4(q^2)}{(q^2-M_3^2(q^2)+i\ep)(q^2-M_8^2(q^2)+i\ep)^2}
  +\cdots \nonumber \\
 &=&\frac{i}{q^2-M_8^2(q^2)-\frac{M_{38}^4(q^2)}{q^2-M_3^2(q^2)}+i\ep}.
\end{eqnarray}
Then the masses of $\pi^0$ and $\eta_8$ are just solutions of the
following equations,
\begin{eqnarray}\label{5.7}
m_{\pi^0}^2&=&M_3^2(m_{\pi^0}^2)+\frac{M_{38}^2(m_{\pi^0}^2)}
   {m_{\pi^0}^2-M_8^2(m_{\pi^0}^2)}, \nonumber \\
m_{\eta^8}^2&=&M_8^2(m_{\eta_8}^2)+\frac{M_{38}^2(m_{\eta_8}^2)} 
   {m_{\eta_8}^2-M_8^2(m_{\eta_8}^2)}.   
\end{eqnarray} 

In eqs.(~\ref{3.16}) and (~\ref{3.23}), the parameters $\kappa$, $F_0$ and
$B_0$ are still not determined. In order to determine them and three light
quark masses, six inputs are required. In this paper we choose
$f_{\pi^+}=185.2\pm 0.5$MeV, $f_{_{K^+}}=226.0\pm 2.5$MeV,
$m_{\pi^0}=134.98$MeV, $m_{_{K^0}}=497.67$MeV and $(m_{_{K^+}})_{\rm
QCD}=491.6\pm 0.1$MeV. Another input is $m_d-m_u=3.9\pm 0.22$MeV, which
is extracted from $\omega\rightarrow\pi^+\pi^-$ decay at energy scale
$\mu=m_\rho$ in ref.\cite{Omega}. Recalling $m=480$MeV, $g_A=0.75$ and
$g=\pi^{-1}$ for $N_c=3$, we can fit light quark masses as in table 1.

In table 1, the errors of results are from uncertainties in decay constant
of $K^+$. electromagnetic mass splitting of $K$-mesons and isospin
violation parameter $m_d-m_u$ respectively. The first column in table 1 we
show the results for $f_{_{K^+}}=223.5$MeV.  The third column corresponds
the center value of $f_{_{K^+}}$, $226.0$MeV and the fifth column
corresponds $f_{_{K^+}}=228.5$MeV. Then from table 1 we have
\begin{eqnarray}\label{5.8}
m_s&=&160\pm 15.5{\rm MeV},\hspace{0.5in}
m_d=7.9\pm 2.7{\rm MeV},\hspace{0.5in}
m_u=4.1\pm 1.5{\rm MeV},\nonumber \\
\frac{m_s}{m_d}&=&20.2\pm 3.0,\hspace{0.8in}
\frac{m_u}{m_d}=0.5\pm 0.09.
\end{eqnarray}
Here the large errors are from the uncertainty of $f_{_{K^+}}$. From table
1 we also have $(m_{\pi^+}-m_{\pi^0})_{\rm QCD}=-0.25$MeV(in this paper 
the contribution from $\pi_3-\eta'$ mixing is neglected). This result
allows that the electromagnetic mass splitting of pion is
$(m_{\pi^+}-m_{\pi^0})_{\rm e.m.}=4.8$MeV.
\begin{table}[hpb]
\centering
 \begin{tabular}{cccccc}  
        &Fit 1&Fit 2&Fit 3& Fit 4&Fit 5 \\ \hline
$^{\dag}f_{\pi^+}$&185.2&185.2&185.2&185.2&185.2 \\
$^{\dag}f_{_{K^+}}$&223.5&224.1&226.0&227.8&228.5 \\
$^{\dag}m_{\pi^0}$&134.98&134.98&134.98&134.98&134.98 \\
$^{\dag}m_{_{K^0}}$&497.67&497.67&497.67&497.67&497.67 \\
$^{\dag}(m_{_{K^+}})_{\rm QCD}$&$491.6\pm 0.1$&$491.6\pm 0.1$&
  $491.6\pm 0.1$&$491.6\pm 0.1$&$491.6\pm 0.1$ \\
$^{\dag}B(\omega\rightarrow\pi\pi)$&$1.95\%$&$1.95\%$&$(2.11\pm 0.20)\%$
  &$(2.21\pm 0.25)\%$&$(2.21\pm 0.30)\%$ \\ 
$m_s$&144.2&150.0&161.8 &172.0&175.4 \\
$m_d$&$6.26$&$6.83$&$7.94\pm 0.07$&$8.94\pm 0.1$&
      $9.53\pm 0.11$ \\
$m_u$&$2.56$&$3.13$&$4.10\mp 0.07$&$5.03\mp 0.1$&
      $5.63\mp 0.11$ \\
$m_u+m_d$&8.92&9.96&12.04&13.97&15.16 \\
$m_d-m_u$&$3.7$&$3.7$&$3.84\pm 0.14$&$3.9\pm 0.2$&$3.9\pm 0.22$ \\
$f_{_{K^0}}$&224.0&224.7&226.6&228.4&229.0 \\
$(m_{\pi^+})_{\rm QCD}$&134.74&134.74&134.73&134.73&134.73 \\
$m_{\eta_8}$&608.8&616.9&628.8&636.8&641.0 \\
$\kappa$&0.5&0.4&0.3&0.25&0.2 \\
$F_0$&156.95&156.7&156.1&155.6&155.3 \\
$B_0$&2141.1&1886.6&1551.1&1328.6&1216.8 \\
\end{tabular}
\begin{minipage}{5in}
\caption{Light current quark masses predicted by the masses and decay
constants of pseudoscalar mesons at energy scale $\mu=m_\rho$. Here
$\kappa$ is dimensionless, other dimensionful quantities are in MeV, and
$\dag$ denotes input. The formula for branching ratio of
$\omega\rightarrow\pi^+\pi^-$ can be found in ref.[15]. It
determines the isospin breaking patameter $m_d-m_u$.}
\end{minipage}
\end{table}

\section{Discussion and Summary}

In this paper we study information on light quark masses at energy scale
$\mu=m_\rho$ in the framework of chiral constituent quark model. The
analysis of quark masses beyond the next to leading order in the chiral
expansion is a challenging subject. The attempt is stoped in ChPT due to
the difficulties mentioned in Introduction. In approach of quark
model, however, it is not necessary to treat the light current quark
masses as the quantities used to construct a perturbative expansion. In
addtion, since the light current quark masses can be defined uniquely, 
the Kaplan-Manohar ambiguity is avoided in this formalism. From
eq.~(\ref{5.8}) $m_u\neq 0$ is also confirmed in our results. This
conclusion supports viewpoint of ref.\cite{Leut90,Donoghue92} but disagree
with one of ref.\cite{Choi88}.

Our results are yielded by a non-perturbative method and contain complete
information on light quark masses. Not only quark mass ratios but also
individual light quark masses are obtained. The results agree with one
obtained by other approachs(e.g., QCD sum rule or ChPT) well.

Finally, it is interesting to expand non-perturtation results (\ref{5.1})
up to the next to leading order of the chiral expansion and compare with
one of ChPT. If we neglect the mass difference $m_d-m_u$, up to this order
the masses of decay constants of koan and pion read
\begin{eqnarray}\label{6.1}
m_\pi^2&=&B_0(m_u+m_d)\{1+\frac{1}{2}\mu_\pi-\frac{1}{6}\mu_{\eta_8}
   +[\frac{1}{2}(3-\kappa^2)+\frac{3m^2}{\pi^2F_0^2}
     (\frac{m}{B_0}-\kappa g_A-\frac{g_A^2}{2}-\frac{B_0}{6m}g_A^2)]  
   \frac{m_u+m_d}{m}\}, \nonumber \\
m_{_K}^2&=&B_0(\hat{m}+m_s)\{1+\frac{1}{3}\mu_{\eta_8}
   +[\frac{1}{2}(3-\kappa^2)+\frac{3m^2}{\pi^2F_0^2} 
     (\frac{m}{B_0}-\kappa g_A-\frac{g_A^2}{2}-\frac{B_0}{6m}g_A^2)]
   \frac{\hat{m}+m_s}{m}\}, \\
f_\pi&=&F_0\{1-\mu_\pi-\frac{1}{2}\mu_{_K}+\frac{3}{2\pi^2}g_A^2
   \frac{m(m_u+m_d)}{F_0^2}\}, \nonumber \\
f_{_K}&=&F_0\{1-\frac{3}{8}\mu_\pi-\frac{3}{4}\mu_{_K}
   -\frac{3}{8}\mu_{\eta_8}+\frac{3}{2\pi^2}g_A^2
   \frac{m(\hat{m}+m_s)}{F_0^2}\}. \nonumber 
\end{eqnarray}
Comparing the above equation with one of ChPT\cite{GL85a}, we predict the
$O(p^4)$ chiral coupling constants, $L_4,\;L_5\;L_6$ and $L_8$, as follow
\begin{eqnarray}\label{6.2}
L_4&=&L_6=0, \hspace{1in}L_5=\frac{3m}{32\pi^2B_0}g_A^2, \nonumber \\
L_8&=&\frac{F_0^2}{128B_0m}(3-\kappa^2)+\frac{3m}{64\pi^2B_0}
  (\frac{m}{B_0}-\kappa g_A-\frac{g_A^2}{2}-\frac{B_0}{6m}g_A^2)
  +\frac{L_5}{2}.
\end{eqnarray}
Numerically, inuptting $\kappa=0.35\pm 0.15$, $F_0=0.156$GeV and
$B_0=1.6\pm 0.4$GeV, we obtain $L_5=(1.6\pm 0.3)\times 10^{-3}$ and   
$L_8=(0.7\pm 0.5)\times 10^{-3}$. These value well agree with one of ChPT,
$L_5=(1.4\pm 0.5)\times 10^{-3}$ and $L_8=(0.9\pm 0.3)\times 10^{-3}$ at
energy scale $\mu=m_\rho$. This fact can interpret why the light quark
mass ratio in eq.~(\ref{5.8}) are close to results extracted from ChPT by
Leutwyler\cite{Leut96}. The above results indicate that the contribution
from scalar meson resonance exchange is small. In fact, in hardon
spectrum, there is no scalar meson otect or singnlet which belong to
composited fields of $q\bar{q}$. Thus it is a {\it ad hoc} assumption to
agrue some low energy coupling constants, such as $L_5$ and $L_8$,
receiving large contribution from scalar meson exchange.

\end{document}